\documentclass[aps,pre,twocolumn,showpacs]{revtex4-1}
\usepackage{graphicx}
\usepackage{dcolumn}
\usepackage{bm}
\usepackage{amsmath}
\usepackage{mathrsfs}
\begin{document}

\title{Simple model of fractal networks formed by self-organized critical dynamics}

\author{Shogo Mizutaka}
\email{shogo.mizutaka.sci@vc.ibaraki.ac.jp}
\affiliation{Department of Mathematics and Informatics, Ibaraki University, 2-1-1 Bunkyo, Mito, Japan 310-8512}

\date{\today}
\begin{abstract}
In this paper, a simple dynamical model in which fractal networks are formed by
self-organized critical (SOC) dynamics is proposed; the proposed model
consists of growth and collapse processes. It has been shown that SOC
dynamics are realized by the combined processes in the model. Thus, the distributions
of the cluster size and collapse size follow a power-law function in the
stationary state. Moreover, through SOC dynamics, the networks become
fractal in nature. The criticality of SOC dynamics is the same as the universality class
of mean-field theory. The model explains the possibility that the fractal nature
in complex networks emerges by SOC dynamics in a manner similar to the case with
fractal objects embedded in a Euclidean space.
\end{abstract}


\maketitle

\section{Introduction}
Networks consisting of elements and their interactions are observed
over a wide range from nature to society. Most large-scale complex
networks share some common structural properties, including the
scale-free property, small-world or fractal nature, degree--degree
correlation, and community structures \cite{Albert02,Dorogovtsev02}.
Based on the relation between the average path length
and system size (i.e., the total number of nodes in a network), real-world
networks are classified into two types, namely small-world and fractal networks. First, in the case of
small-world networks, the distance between any two nodes is
extremely small compared with the system size. More precisely, in small-world networks,
the average path length increases only with the logarithm of the system
size at most. In other words, if we cover a small-world network
using boxes with a linear size of $l_{{\rm B}}$, the relation
\begin{equation}
N_{{\rm B}}(l_{\rm B})\propto e^{-l_{\rm B}/l_{0}}
\label{SW}
\end{equation}
is satisfied. Here, $N_{\rm B}(l_{\rm B})$ represents the minimum number of boxes required
to cover a given network, while $l_{0}$ is the characteristic length
of the network.
Second, in contrast, fractal networks satisfy the following relation:
\begin{equation}
N_{\rm B}(l_{\rm B})\propto l_{\rm B}^{-d_{\rm B}},
\label{fractal}
\end{equation}
where $d_{\rm B}~(<\infty)$ represents the fractal dimension. It is evident that Eq.~(\ref{fractal})
does not include a characteristic length.
Because the small-world nature expressed in Eq.~(\ref{SW}) corresponds
to the case of $d_{\rm B} \to \infty$ in Eq.~(\ref{fractal}),
these two concepts are conflicting.

Although it is known that the small-world structure in a network
is formed by the generation of shortcut edges between long-range
nodes \cite{Watts98}, the origin of the fractality in
networks is not yet understood well \cite{Fujiki17,Watanabe15,Yook05}.
However, it is known that most of the fractal objects embedded in a
Euclidean space, such as coastlines and the branches of trees, emerge as a
consequence that a dynamical system is spontaneously driven toward a
critical point as an attractor and fluctuates near it
\cite{Bak87,Bak93,Paczuski96,Markovic14,Drossel92,Takayasu92,Rinaldo93,Sapoval04,deArcangelis02}.
We call the phenomenon {\textit {self-organized criticality}}. In
self-organized critical (SOC) dynamics, the characteristic scales of various
quantities vanish, and their distributions obey a power-law function.

It is natural to expect that the fractal structures of networks are formed
by SOC dynamics as is the case with fractal objects in a Euclidean space.
In fact, it has recently been shown that the fractality of networks
is formed by SOC dynamics \cite{Watanabe15}. The basic idea of the model
proposed in Ref.~\cite{Watanabe15} is to combine the growth of a network
and its collapse by cascading overload failures, as argued in
Ref.~\cite{Mizutaka15,Mizutaka17}. In particular, the combined dynamics
of the two percolation processes drive the network into a critical state;
consequently, the network structure exhibits fractality in the stationary
state. Moreover, the distribution of various quantities such as the cluster size,
cascade size, and waiting time between two adjacent cascades obeys the power-law
form. In the model, however, the criticality of the SOC dynamics is not clear
owing to the complication of the model, which contains many predetermined parameters.
Thus, this complication does not allow us to understand what factors change the
criticality and what universality class is realized. It is important to
clarify these points to better understand the fractal nature observed in networks and
to establish fundamental knowledge in the field of network science.
Therefore, in this paper, we propose a simple dynamical model combining \textit{growth}
and \textit{collapse} processes in the network.
We apply the Erd\H{o}s-R\'{e}nyi process \cite{Erdos60} and cascading failure process
based on the threshold model proposed by Watts \cite{Watts02} to the growth and collapse
processes, respectively. Our main result shows that SOC dynamics
emerge by the combined dynamics of the growth and collapse processes, and
the networks become fractal in the stationary state.

The rest of this paper is organized as follows. In Sec.~\ref{sec:model},
we briefly review the threshold model. Then, we introduce our proposed model.
The results and their discussion are presented in Sec.~\ref{sec:result}. Finally,
Sec.~\ref{sec:conclusion} includes our concluding remarks.

\section{Model}
\label{sec:model}
\subsection{Collapse Process: Threshold Model}
\label{sec:pre_model}
Since we apply the threshold model proposed by Watts \cite{Watts02}
to a collapse process in the model proposed in this paper, we briefly review the
outline of this model in this subsection. It is noted that the cascading failure in the threshold
model spreads along the nearest-neighbor nodes, which imitates the spread of rumors in the
case of social networks and those of chain bankruptcies in financial networks, among others.
The manner of the threshold model is as follows:

(i) Construct a network with $N$ nodes. Every node can only be in one of the two states (either active or inactive). All nodes are in the inactive state and have a threshold $\phi$.

(ii) Change the state of a randomly selected node to the active state.

(iii) Change the state of a node to be active if the ratio
$\Phi_i$ of the number of active adjacent nodes to the degree
of the selected node is larger than a given threshold value $\phi$.
Repeat this update iteratively until no new active nodes are
generated (i.e., a cascading process).

The threshold $\phi$ represents the amount of support a particular node needs from
its nearest neighbors. The cascade size is defined
as the number of active nodes after the cascading process. By
demonstrating the relation between the threshold value $\phi$ and
the characteristics of the network structures, such as the degree
distribution and average degree, it has been clarified that
a single activation of a node would cause a global cascade that
spreads throughout networks under certain conditions \cite{Watts02}.
In particular, for Erd\H{o}s-R\'{e}nyi random graphs, a phase diagram showing whether a
global cascade occurs can be obtained using the generating function
method for a plane formed by the average degree $\langle k\rangle$ and threshold value
$\phi$. The previous results in Ref.~\cite{Watts02} show that a phase
border exists for $\langle k\rangle =1$ and $\phi < 1/4$.

\subsection{Present Model}
In this paper, a dynamical model, which is the threshold model
with a growth process, is proposed. The evolutional manner of the
present model is as follows:

(i) Prepare $N$ isolated nodes that can be only be in one of two states
(either the active or inactive state). Each node is in the inactive state with a threshold
$\phi$.

(ii) At each time step $t$, connect two randomly selected nodes that
are not yet connected. Then, every node in the system changes its
state to the active state with the probability $p$. If there exist
active nodes in the system, go to Step (iii). Otherwise,
repeat Step (ii).

(iii) Change the state of a node to active if the ratio
$\Phi_i$ of the number of active adjacent nodes to the degree
of the node is larger than the given threshold value $\phi$.
Repeat this update iteratively until no new active nodes are
generated. After completing the iterative renewals of node
states, remove the edges of all active nodes and restore the
states of these nodes to the inactive state.

(iv) Repeat Steps (ii)--(iii).

It is noted that the growth process is the evolution of
random graphs in the Erd\H{o}s-R\'{e}nyi model \cite{Erdos60}.
The inverse of the probability $p$ would be the average lifetime
of nodes if the collapse processes never occurred in the threshold model.
For such an instability of each node, the network cannot reach
a complete graph, in which every pair of nodes is connected.

\begin{figure}[t!]
\begin{center}
\includegraphics[width=0.45\textwidth]{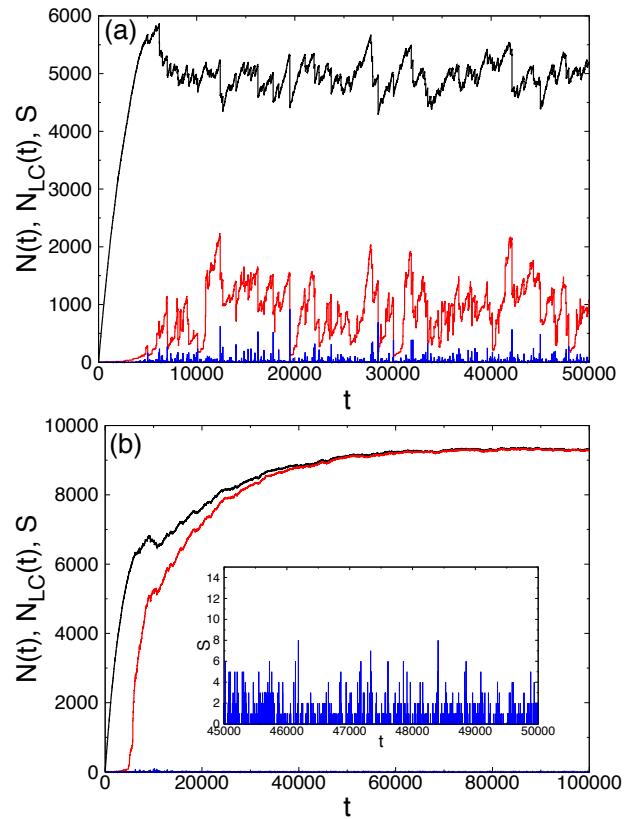}
\caption{Time dependencies of the total number $N(t)$ (top, black line)
of nodes excluding isolated nodes, the size $N_{\text LC}(t)$
(middle, red line) of the largest component, and the cascade size
$S$ (bottom, blue spiked line) for (a) $\phi=0.2$ and (b) $\phi=0.3$.
The inset of panel (b) shows a magnified illustration that depicts the
time dependence of the cascade size $S$. The results are obtained
under the conditions $N=10^{4}$ and $p=10^{-5}$.}
\label{fig1}
\end{center}
\end{figure}

\section{Results \& Discussion}
\label{sec:result}
\begin{figure}[t!]
\begin{center}
\includegraphics[width=0.45\textwidth]{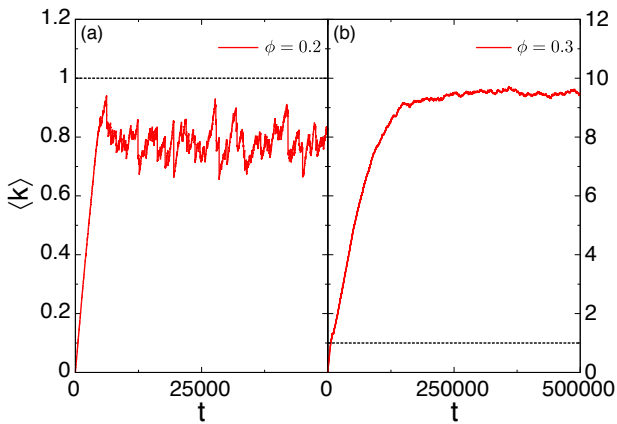}
\caption{Time dependence of the average degree $\langle k\rangle$ for
(a) $\phi=0.2$ and (b) $\phi=0.3$. The horizontal dashed lines in both
panels are for $\langle k\rangle=1$, which represents the condition of the percolation
transition in the Erd\H{o}s-R\'{e}nyi random graph. The results are obtained
under the conditions $N=10^{4}$ and $p=10^{-5}$.
}
\label{fig2}
\end{center}
\end{figure}

It is interesting to clarify several properties of networks in
the stationary state. Figure~\ref{fig1} shows the time evolution of the
network size $N(t)$ excluding isolated nodes, the size $N_{\rm LC}(t)$
of the largest cluster, and the cascade size $S$ defined
as the number of active nodes after completing the cascading process
at each time step $t$.
As seen from both panels in Fig.~\ref{fig1}, $N(t)$ increases almost
monotonically with the time $t$ at the early stage and then attains the
stationary state. However, the behaviors of several quantities in the stationary state in the
two panels are entirely different. In the top panel corresponding to the result
for $\phi=0.2$, each quantity largely fluctuates.
In contrast, for $\phi=0.3$, the sizes $N(t)$, $N_{\rm LC}(t)$,
and $S$ do not largely fluctuate, and the nodes with edges almost form a singly connected component.
Furthermore, in Fig.~\ref{fig2}, the time dependencies of the average degree for $\phi=0.2$
and $\phi=0.3$ are shown. In this figure, the average degree $\langle k\rangle$ at time $t$
is defined as the ratio of double the number of edges to the total number of nodes
$N$ in the system. For $\phi=0.2$, the average degree fluctuates near
$\langle k\rangle=1$ after reaching the stationary state, while the average degree for
$\phi=0.3$ fluctuates near $\langle k\rangle=10$.
These differences imply that the behaviors of the
cluster-size distributions for $\phi=0.2$ and $\phi=0.3$ change drastically.

\begin{figure}[t!]
\begin{center}
\includegraphics[width=0.45\textwidth]{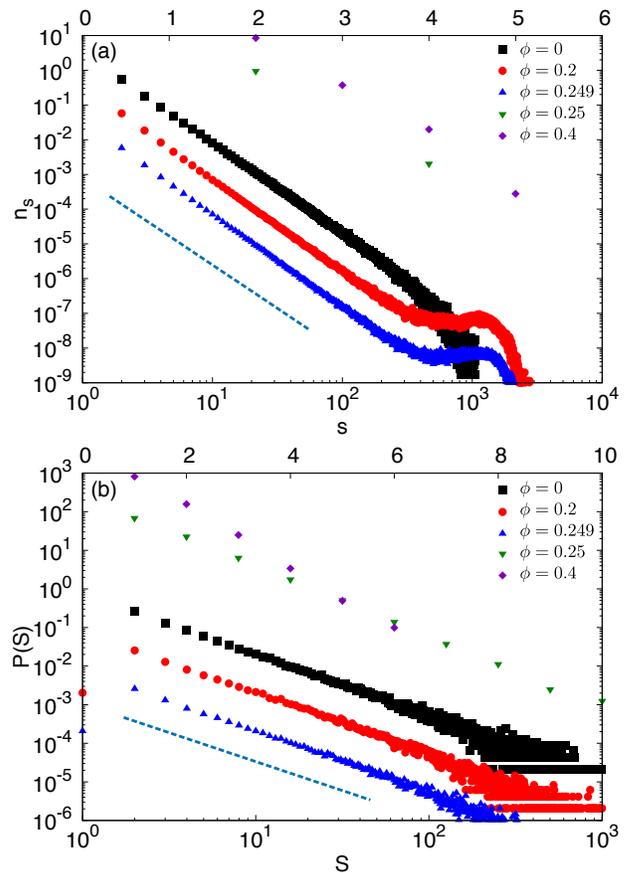}
\caption {Distribution functions of (a) $n_{s}$ of the cluster size $s$ and (b) $P(S)$ of the avalanche size $S$ for various values of
the threshold $\phi$. Although the results for $\phi<1/4$ are plotted on a double-logarithmic
scale (left and lower axes), the results for $\phi\ge 1/4$ are plotted on a
semilogarithmic scale (left and upper axes) in both panels.
The distributions $n_{s}$ and $P(S)$ are obtained from the data until
$t=10^{6}$ after attaining the stationary state.
The plots are vertically shifted for clarity. The slopes of the guide lines
(dashed light-blue lines) are $5/3$ and $3/2$ in (a) and (b), respectively.
The other parameters are set as follows: $N=10^{4}$ and $p=10^{-5}$.
}
\label{fig3}
\end{center}
\end{figure}

It is important to ascertain whether SOC dynamics emerge
in the behaviors depicted in Fig.~\ref{fig1}. Thus, in order to confirm this,
 we numerically investigate the distribution functions
$n_{s}$ and $P(S)$ of the cluster size $s$ and cascade
size $S$, respectively.
The top panel in Fig.~\ref{fig3} shows the distribution $n_{s}$ of
the cluster size $s$ averaged over the system after it has attained the stationary
state for various values of the threshold $\phi$. For $\phi<1/4$, the distribution $n_{s}$
obeys the following power law:
\begin{equation}
n_{s}\sim s^{-\tau}.
\end{equation}
This implies that networks remain near the critical point of the percolation
transition by the combined dynamics. Moreover, the slope of each result
for $\phi < 1/4$ is parallel to the guide line (dashed light-blue line),
which indicates the critical exponent $\tau=5/2$ for $\phi < 1/4$.
This result indicates that the exponent $\tau$ does not depend on
the threshold $\phi$. In addition, the value of the exponent corresponds to
that of the exponent characterizing the cluster distribution in the Erd\H{o}s-R\'{e}yni
random graph at criticality. Furthermore, the behavior of the average degree
depicted in Fig.~\ref{fig2}~(a) implies that a network formed in the present model attains
the critical point in the Erd\H{o}s-R\'{e}nyi random graph.
In contrast, for $\phi\ge 1/4$, the distribution $n_{s}$ decays
exponentially, which indicates that below the threshold $\phi_{\rm ch}=1/4$,
the distribution of $n_{s}$ obeys the abovementioned power law,
which is consistent with the previous result for the threshold model \cite{Watts02}
described in Sec.~\ref{sec:pre_model}.

In the bottom panel of Fig.~\ref{fig3}, the distribution $P(S)$ of the
cascade size $S$ is shown. The behavior of the distribution $P(S)$
of the avalanche size for $\phi$ is similar to that of $n_{s}$.
For $\phi<1/4$, the distribution $P(S)$ also follows a power-law form,
\begin{equation}
P(S)\sim S^{-\alpha},
\end{equation}
with the exponent $\alpha=3/2$.
The behavior of the dynamics in the present model is similar to that in
the forest fire model proposed by Drossel and Schwabl \cite{Drossel92}.
The activation rate $p$ of each node and the addition of a new edge for each time step
in our model correspond to the lightning probability and growing probability
of a tree in the forest fire model, respectively. In particular, in the case of $\phi=0$, our model is equivalent to a random neighbor treatment for the forest fire model \cite{Christensen93}.
\begin{figure}[t!]
\begin{center}
\includegraphics[width=0.45\textwidth]{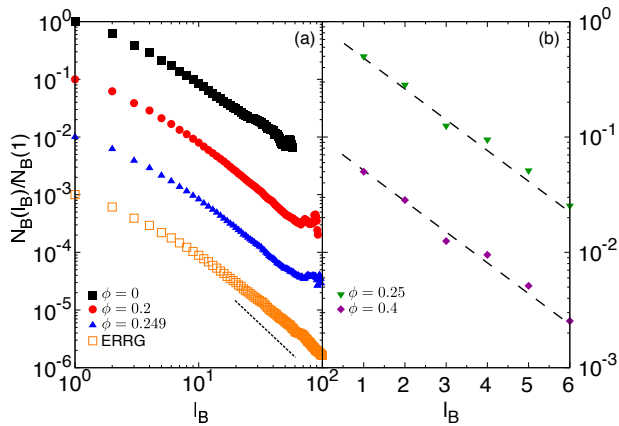}
\caption{$N_{\rm B}(l_{\rm B})$ for the giant components in networks in the
stationary state under the conditions (a) $\phi=0$, $0.2$, and $0.249$
and (b) $\phi=0.25$ and $0.4$. The longitudinal axis indicates
$N_{\rm B}(l_{\rm B})/N_{\rm B}(1)$ averaged over $10^3$ realizations of networks. In panel
(a), the plots present the results for $\phi=0$, $0.2$, and $0.249$ from top to bottom, respectively.
The bottom-most plot shows the results for the giant components of
an Erd\H{o}s-R\'{e}nyi random graph with $N=10^4$ and $\langle k\rangle=1.0$,
i.e., at criticality. The dashed line with a slope of $-2$ represents the
fractal dimension of the Erd\H{o}s-R\'{e}nyi random graph.
}
\label{fig4}
\end{center}
\end{figure}

The reason why the exponent $\alpha=3/2$ is as follows.
A node in a cluster changes a state from inactive to active with the probability
$p$. Then, the expectation of the occurrence of a state change with the probability $p$ in
an $s$-cluster is proportional to $psn_{s}$. Because the size of the cascade is equal
to the size of cluster for $\phi=0$, the exponents $\tau$ and $\alpha$ satisfy the relation
$\alpha= \tau-1$. This relation holds even for $\phi=0.2$
and $0.249$ if the system is sufficiently large.

Finally, we study the properties of network structures in the stationary
state. If the network has a fractal nature, the relation defined by Eq.~(\ref{fractal})
is satisfied.
We estimate the minimum number of subgraphs $N_{\rm B}(l_{\rm B})$ for the largest components
using the compact-box-burning algorithm \cite{Song07}. Figures~\ref{fig4}(a) and (b) show the linear size $l_{\rm B}$ dependence of $N_{\rm B}(l_{\rm B})/N_{\rm B}(1)$
for different values of the threshold $\phi$, where $N_{\rm B}(1)$ is
the number of nodes belonging to the largest component. In panel (a),
we also plot the results of a fractal analysis for the giant component of
Erd\H{o}s-R\'{e}nyi random graphs with the average degree
$\langle k\rangle=1$, i.e., at criticality. As shown in Fig.~\ref{fig4},
$N_{\rm B}(l_{\rm B})$ for $\phi<1/4$ satisfies the relation given by Eq.~(\ref{fractal}).
This result indicates that the fractality in networks appears by
SOC dynamics in a manner analogous with the case of a Euclidean space.
In addition, as seen from Fig.~\ref{fig4}(a), in the plots for
the present model with $\phi<1/4$ and Erd\H{o}s-R\'{e}nyi random model,
the tails are parallel, which indicates that the fractality of networks
with $\phi<1/4$ corresponds to that of the Erd\H{o}s-R\'{e}nyi random graphs at criticality.
The slightly difference between the analysis (dashed guide line) and present simulated
results is based on the finite-size effect that the simulated results experience.
Therefore, we conclude that networks governed by the present
SOC dynamics are driven towards the percolation transition point for
Erd\H{o}s-R\'{e}nyi random graphs, which, in turn, indicates that the criticality of the
present model is the same as the universality class of mean-field theory.

\section{Conclusion}
\label{sec:conclusion}
In this paper, we proposed a simple dynamical model in which
fractal networks are formed by SOC dynamics;
in particular, our model consists of the Erd\H{o}s-R\'{e}nyi process \cite{Erdos60}
and a cascading failure process based on the threshold model \cite{Watts02}. In the proposed
model, the distributions of the cluster size and cascade size follow a power-law
function in the stationary state. Moreover, through SOC dynamics, networks in the
stationary state become fractal. Thus, this indicates that
fractality in networks would emerge by SOC dynamics, as is the case with
fractal objects embedded in a Euclidean space.
In addition, our simulation results show that the universality class of
the self-organized criticality in the developed model corresponds to that of
the Erd\H{o}s-R\'{e}nyi random graph. In particular, for
the model parameter $\phi=0$, the present model
corresponds to the mean-field treatment for the forest fire model \cite{Christensen93}.

The combined dynamics realizing SOC dynamics do not have to be the
Erd\H{o}s-R\'{e}nyi process and the cascading failure process based on the
threshold model. The SOC dynamics leading to fractality in networks would be realized
in different scenarios. For instance, a simple and possible candidate of the collapse process
is the spread of diseases based on the susceptible–infected–recovered model. Considering that the universality class of the SOC dynamics in the
present model is the same as that of the Erd\H{o}s-R\'{e}nyi random graph,
the universality class might be related to a percolation process, i.e., a growth process.
In order to clarify the diversity of fractal networks, it is important to
understand whether growth processes change the universality class.

Furthermore, most of the fractal networks observed in the real world, such
as the World Wide Web or metabolic networks, also have the scale-free property \cite{Yook05}.
Recent work has highlighted that such networks possess long-range repulsive correlations in
nodes with similar degrees \cite{Fujiki17}. It is interesting to clarify whether
fractal networks with the scale-free property are formed by SOC dynamics,
which explain the long-range degree correlations if the former is true.

\begin{acknowledgements}
The author would like to thank K. Yakubo for fruitful discussions. This work was supported by
a Grant-in-Aid for Early-Career Scientists (No.~18K13473) and a Grant-in-Aid for JSPS Research
Fellows (No.~18J00527) from the Japan Society for the Promotion of Science.
\end{acknowledgements}

\end{document}